\documentclass[traditabstract]{aa}  
\usepackage{graphicx,natbib,psfig,amssym}

\def\lapp{\mathbin{\raise2pt \hbox{$<$} \hskip-9pt \lower4pt \hbox{$\sim$}}}
\def\gapp{\mathbin{\raise2pt \hbox{$>$} \hskip-9pt \lower4pt \hbox{$\sim$}}}
\newcommand{\ltsima} {$\; \buildrel < \over \sim \;$} 
\newcommand{\gtsima} {$\; \buildrel > \over \sim \;$} 
\newcommand{\lta} {\lower.5ex\hbox{\ltsima}} 
\newcommand{\gta} {\lower.5ex\hbox{\gtsima}}

\newcommand{\ergs}{\>{\rm erg}\,{\rm s}^{-1}}
\newcommand{\ergsHz}{\>{\rm erg}\,{\rm s}^{-1}\,{\rm Hz}^{-1}}

\newcommand{\kms}{$\rm{\,km \,s}^{-1}$}

\begin{document}
\title{Accretion and nuclear activity in Virgo early-type galaxies.}
\subtitle{}
  \titlerunning{Accretion and nuclear activity in Virgo early-type galaxies.}
\authorrunning{Vattakunnel et al.}

\author{Shaji Vattakunnel \inst{1,3}, Edoardo Trussoni\inst{2}, Alessandro
  Capetti\inst{2}, \and Ranieri D. Baldi\inst{2,3}} \offprints{S. Vattakunnel}
\institute{Dipartimento di Fisica Universit\`a di Trieste, piazzale Europa 1,
  I-34127, Trieste, Italy \and INAF - Osservatorio Astronomico di Torino, Via
  Osservatorio 20, I-10025 Pino Torinese, Italy \and
  Universit\`a di Torino, via P. Giuria 1, 10125 Torino, Italy\\
  \email{vattakunnel@oats.inaf.it,trussoni@oato.inaf.it,capetti@oato.inaf.it,
    baldi@oato.inaf.it}} 

\date{}
 
\abstract{We use Chandra observations to estimate the accretion rate of hot
  gas onto the central supermassive black hole in four giant (of stellar mass
  $M_* \sim 10^{11} - 10^{12} M_{\sun}$) early-type galaxies located in the
  Virgo cluster. They are characterized by an extremely low radio luminosity,
  in the range $L \lesssim3\times10^{25} - 10^{27} \ergsHz$. We find that,
  accordingly, accretion in these objects occurs at an extremely low rate,
  $0.2 - 3.7 \times 10^{-3} M_{\odot}$ yr$^{-1}$, and that they smoothly
  extend the relation accretion - jet power found for more powerful
  radio-galaxies. This confirms the dominant role of hot gas and of the
  galactic coronae in powering radio-loud active galactic nuclei across $\sim$
  4 orders of magnitude in luminosity. A suggestive trend between jet power
  and location within the cluster also emerges.}


\maketitle

\section{Introduction}
\label{intro}

The origin of the energetic processes occurring in active galactic nuclei
(AGN) has been one of the most relevant astrophysical problems in the last
decades. It is well established that the main driving mechanism is the release
of gravitational energy of flows accreting on supermassive black holes (SMBH)
for both radio-loud and radio-quiet AGN. While in the latter class the energy
is mainly released into electromagnetic radiation, in radio-loud AGN a
substantial fraction of the accretion power is converted into the kinetic
  power of relativistic jets. What is still a matter of debate is the origin of
the accreting material that fuels the AGN, how this is channeled onto the
SMBH, and whether and how it is eventually accelerated into a
relativistic collimated outflow.

\begin{table*}
 \caption{Properties of the sample of core galaxies in the VCC}
\begin{tabular}{|c|c|c|c|c|c|c|c|c|}
  \hline
  Source   & Alt. names      & $z$ &  $D$ &  sub cl.    & $d$ & $V_z$ & Log
  $L_{\rm core}$ &  Log $M_{BH}$    \\
  \hline  
  VCC~~731 & NGC~4365       & 0.00415  & 23.3          &  B & 6.2  &  246 &$<$ 25.84 & 8.59    \\
  VCC~~798 & M~85, NGC~4382 & 0.00243  & 17.9          &  A & 1.8  &  578 &$<$ 25.60 & 7.93    \\
  VCC~~881 & M~86, NGC~4406 & -0.00081 & 17.9          &  A & 0.7  & 1584 &    25.94 & 8.52   \\
  VCC~1535 & NGC~4526       & 0.00149  & 16.5$^{\rm a}$ &  B &$>1.1$&  549 &    27.03 & 8.59   \\
  \hline                                                                                                  
  VCC~~763 & M~84, NGC~4374 & 0.00354  & 18.4          &  A & 1.2  &  247 &    28.87 & 9.00$^{+0.48}_{-0.40}$ \\
  VCC~1226 & M~49, NGC~4472 & 0.00333  & 17.1          &  B dom. & -    &    - &    27.10 & 8.78  \\
  VCC~1316 & M~87, NGC~4486 & 0.00436  & 17.2          &  A dom.& -    &    - &    29.99 & 9.53$^{+0.12}_{-0.15}$  \\
  VCC~1632 & M~89, NGC~4552 & 0.00113  & 16.2          &  A & 1.0  &  967 &    28.24 & 8.54 \\ 
  VCC~1939 & NGC~4636       & 0.00313  & 15.6$^{\rm b}$ &  B & 2.7  &   59 &    27.02 & 8.16   \\
  VCC~1978 & M~60, NGC~4649 & 0.00373  & 17.3          &  A & 0.9  &  190 &    27.82 & 9.30 $^{+0.08}_{-0.15}$ \\
  \hline
\end{tabular} 

\medskip
Column description: 1) name; 2)
alternative name; 3) spectroscopic redshift from the NED database; 4) distance in Mpc ($\pm 0.5$
Mpc) from Mei et al. (2007); 
5) sub-cluster to which each galaxy belongs, with the dominant galaxies marked;
6) and 7) three dimensional distance (Mpc) 
and velocity (km s$^{-1}$) along the line of sight 
with respect to the dominant components of the sub-clusters Virgo A or B; 
8) radio core
luminosity ($\ergsHz$) from Capetti et al. (2009); for NGC~4636 we adopted the value of BBC08, adjusted at
the distance assumed in this paper; 9)
mass of the central SMBH in units of $M_{\odot}$ derived from 
the stellar velocity dispersion reported in the HyperLeda database
or from direct measurements as listed by \citet{marconi03}
(when
not reported, we adopt an intrinsic error of $\pm$ 0.23 dex, related to the
dispersion of the $M_{\rm BH}$ - $\sigma_*$ relation). 

\noindent
$^{\rm a}$ For this object there is no distance measurement thus we  
adopted the
average Virgo distance to estimate luminosities;
accordingly its three dimensional distance $d$ from the cluster is a lower limit,
equal to the offset from M~49 in the plane of the sky.

\noindent
$^{\rm b}$ The distance of NGC~4636,
with an associated error of $\pm 0.9$ Mpc,
is from \citet{tonry01}, 
scaled by a factor 1.06, the average ratio between the
distances of \citet{tonry01} and \citet{mei07} for our sample.  
\label{tab1}
\end{table*}

X-ray observations of radio-loud galaxies recently indicated that in these
objects spherical accretion of hot gas can provide the necessary level of
accretion to power the active nucleus. Indeed, \citet{allen06} found a strong
correlation between the jet and the accretion powers in a sample of nine
objects. The jet kinetic energy was estimated from the mechanic work needed to
inflate the cavities observed in their X-ray images.  The same X-ray data
also provided information on the physical parameters of the hot inter-stellar
medium (ISM), yielding estimates of the accretion power exploiting the Bondi
model (1952).

\citeauthor{balmaverde08alias} (2008, here after BBC08) extended Allen's work
considering a larger number of radio-galaxies. Because X-ray cavities are
visible only in a minority of objects, they took advantage of the relation
between jet and radio-core power calibrated by \citet{heinz07}. In this
approximation, the jet kinetic power $P_J$ is estimated directly from the
radio core luminosity, a quantity readily measurable in radio-galaxies. For a
sample of 23 objects with X-ray data of sufficient quality (including the
sources analyzed in Allen et al.) they showed that the correlation between the
jet and accretion power holds over $\sim $ 3 decades, down to a jet power
of $\sim 10^{42}$ $\ergs$. Another result of this correlation is moreover that
on average $\sim 1 \%$ of the available accretion power is converted into jet
kinetic energy, independently of the AGN luminosity. Similar results were
obtained also by \citet{hardcastle07} following a different approach.

Taking advantage of the recent surveys of early-type galaxies in the Virgo
cluster performed in the optical \citep{cote04}, radio \citep{capetti09}, and
X-rays \citep{gallo08} we are in the position of testing the validity of the
correlation accretion/jet powers by including objects of even lower luminosities
than those discussed in the previous studies. We then isolated four giant
galaxies (of stellar mass $M_* \sim 10^{11} - 10^{12} M_{\sun}$)  characterized
by an extremely low radio luminosity (two of these sources are actually
undetected in radio observations), in the range $L \lesssim3\times10^{25} -
10^{27}$ erg s$^{-1}$Hz$^{-1}$, more than 5 orders of magnitude lower than the
brightest objects considered by \citet{allen06} and BBC08.

Furthermore, we have now a total of 10 radio-galaxies located in the Virgo
cluster. It will then be possible to test whether in this environment the
connection between the jet and accretion rates also holds and whether and how
galaxies show different properties, depending on their positions inside the
cluster.

In the following Sect. 2 we outline the main properties of the selected
galaxies, in Sect. 3 we summarize how we estimated, through the X-ray and radio
data, the values of the jet and accretion powers. In Sect. 4 we discuss our
results and their implications for 
the sample considered by BBC08 with the addition of the
four Virgo galaxies examined here. The summary of our
results are reported in Sect. 5. In the Appendix we provide the
details of the analysis of the Chandra X-ray data.

\section{The sample}
\label{sample}

We considered the sample observed by \citet{cote04} for their Hubble Space
Telescope survey of the Virgo cluster (ACS/VCS). They selected a sub-set of
100 early-type galaxies from the Virgo cluster catalog (VCC,
\citealt{binggeli87}) that consists of 2096 galaxies within this $\approx$ 140
deg$^{2}$ region. These data provide us with a detailed analysis of their
optical brightness profiles \citep{ferrarese06}.

\citet{balmaverde06core} found that the nuclear multiwavelength properties
of early-type galaxies are related to their optical brightness profile.  In
particular, radio-loud AGN are found only in `core-galaxies' (hereafter
CoreG), i.e. objects whose brightness profile is better fit by a core-S\'ersic
model compared to  a pure `power-law' S\'ersic model (hereafter
SG) extending to the innermost regions \citep{trujillo04}. 

We here focus on the properties of the sub-sample formed by the nine CoreG
present in the Virgo sample\footnote{The low luminosity VCC~1250 is also
classified as a core-galaxy but this classification is considered as
marginal and should be interpreted with caution because the galaxy morphology
is severely affected by dust patches and a bright nuclear source.}. The
VCC/CoreG are the brightest early-type galaxies in the sample (with the only
pure S\'ersic interloper represented by 9th ranked VCC~1903). This was
expected, considering the well known link between brightness profiles and
luminosity (e.g. \citealt{lauer07}). Because the VCC is complete down to B$_T
\sim 12$ and all CoreG have B$_T$ $ \leq $ 10.78, the VCC/CoreG sub-sample is
also complete. These galaxies can be well compared with the radio-galaxies
studied by \citet{allen06} and BBC08 considering the
similarity of their host properties.

\begin{figure}
\psfig{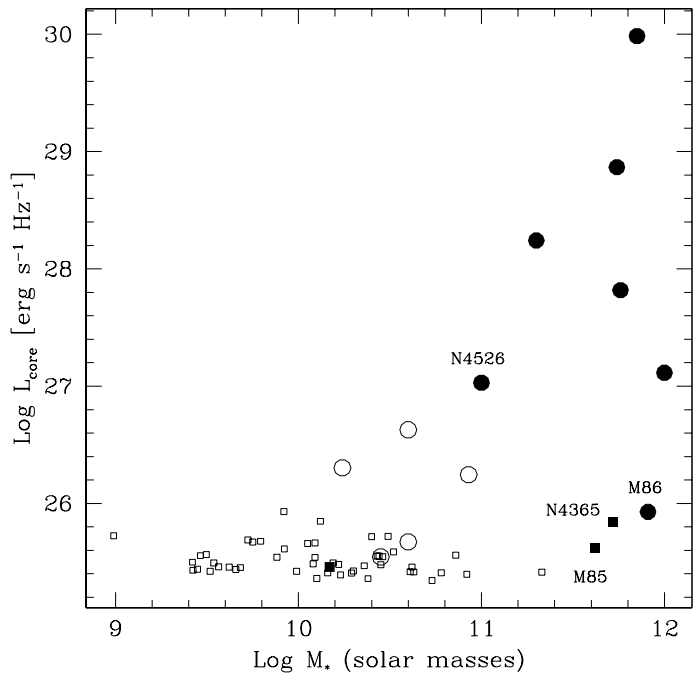}
\caption{Radio core luminosity vs the galaxy stellar masses for the Virgo sample
  analyzed in Capetti et al. (2009). 
  The four galaxies of interest are labeled
  with their Messier or NGC numbers. 
  The radio detected galaxies are marked with large circles, separating them
  on the basis of the optical surface brightness profiles: filled circles are
  CoreG, while empty circles are pure SG. The
  undetected objects are marked with small squares (for clarity, we omit the
  downward pointing arrows), empty for SG, filled for
  CoreG.}
\label{fig1}
\end{figure}

\citet{capetti09} performed a radio survey of the Virgo Cluster, observing
with the Very Large Array (VLA) in A configuration and at 8.4 GHz the 63
early-type galaxies covered by the ACS/VCS, limiting to those brighter with
B$_T$ $ \leq $ 14.4. We here reproduce their Fig. 1, where the nuclear radio
luminosity is plotted against the stellar mass of the galaxy (see
Fig. \ref{fig1}). In agreement with previous studies
(e.g. \citealt{best05b,mauch07}), in the Virgo cluster the detection of a
radio source is also highly favored in more massive galaxies.  The nine
VCC/CoreG have similar masses, all within the range $10^{11} - 10^{12}$
M$_{\sun}$.  Conversely, they show a large scatter in their nuclear radio
luminosity, over more than 4 orders of magnitude. Indeed, while seven are
detected in radio, we could only set an upper limit for the remaining
two. They differ even more in terms of total radio power (measured at 1.4 GHz,
see \citealt{capetti09}), spanning at least over a factor of $\gtrsim 2 \times
10^5$.  This confirms the idea that the relation between radio and optical
luminosity can only be described in terms of a probability distribution.

Below we explore the possibility that the energetic output in these
galaxies is correlated with the accretion rate of the hot, X-ray emitting
gas, in analogy with the results obtained on more powerful sources. Of
particular interest are the four VCC/CoreG\footnote{For easier reference we will
  use the more common Messier or NGC numbers for these sources.} with the
lowest radio
luminosity (namely NGC~4365, M~85, M~86, and NGC~4526), not analyzed in
previous studies, that reach levels as low as $L_{\rm
  core}\lesssim3\times10^{25}$ erg s$^{-1}$Hz$^{-1}$, a further factor of
$\gtrsim 10$ lower than the faintest objects considered in BBC08.

In Table \ref{tab1} we list the main data of the four sources of interest. As we
are also looking for possible connections between the accretion properties and
the cluster environment, we consider the relationships between the galaxies of
our sample and the two main cluster's sub-structures (Virgo A and Virgo B,
with M~87 and M~49 as dominant components, respectively, see
\citealt{binggeli87}). Accordingly we reported also the relative distances $d$
and velocities $V_z$ (along the line of sight) of the various galaxies with
respect to the dominant galaxy of each sub-cluster.

At this stage, we also include in the sample the CoreG VCC~1939 (i.e.
NGC~4636). This source is not covered by the ACS/VCS because it is reported by
\citet{binggeli87} to be a {\sl possible} cluster member, owing to its relative
large angular distance from $\sim 8^{\rm o}$ (2.3 Mpc) from M~49.  However,
its membership appears to be rather secure, considering its distance of
$15.6\pm0.9$ Mpc (compared to D=17.1 Mpc of M~49) and its recession
velocity of only 69 \kms\ relative to M~49.  In Table \ref{tab1} we provide
the data for easy reference also for the six VCC/CoreG already discussed in
previous studies.

\begin{table*}
\caption{ISM properties, accretion rates, jet and accretion powers.} 
\begin{tabular}{|l|c|c|c|c|c|c|c|}
\hline 
Name & $r_B$   & $\alpha$ & $T_B$   & $n_B$  & $\dot M_B$ & log $P_B$ & log $P_j$  \\
\hline
NGC~4365 & 29  & 1.2$^{+0.3}_{-0.2}$ & 0.44 $\pm$ 0.08 & $0.7^{+1.1}_{-0.4}$ & 4.6 & $1.41 \pm 0.38 $ & $<$ -1.68\\
M~85     & 5.5 & 0.8$^{+0.3}_{-0.2}$ & 0.51 $\pm$ 0.23 & $0.7^{+1.0}_{-0.6}$ & 0.2 & $0.05 \pm 0.61$ & $<$ -1.83 \\
M~86     & 17  & 0.4$^{+0.3}_{-0.2}$ & 0.65 $\pm$ 0.02 & $0.2^{+0.4}_{-0.1}$ & 0.5 & $0.45 \pm 0.54$  & -1.61 \\
NGC~4526 & 19  & 0.7$^{+0.3}_{-0.5}$ & 0.68 $\pm$ 0.21 & $0.3^{+1.6}_{-0.2}$ & 1.1 & $0.80 \pm 0.62$  & -0.83 \\
\hline
M~84     & 47 &  0.55$^{+0.19}_{-0.18}$ &  0.71  $\pm$ 0.05 & 0.9$^{+0.6}_{-0.5}$ &   6.6 &  1.58$^{+0.29}_{-0.30}$ & 0.18$^{+0.12}_{-0.16}$ \\
M~49     & 28 & 0.36$^{+0.12}_{-0.12}$ & 0.90 $\pm$ 0.16  & 0.8$^{+0.5}_{-0.3}$  & 12.0   & 1.83$^{+0.23}_{-0.25}$   &  -0.09$^{+0.11}_{-0.15}$    \\
M~87     & 142 & 0.00$^{+0.10}_{-0.10}$ &  0.80 $\pm$ 0.01 & 0.17$^{+0.04}_{-0.03}$   &  24.0   & 2.13$^{+0.40}_{-0.28}$   & 0.53$^{+0.17}_{-0.29}$ \\
M~89     & 17 &  0.83$^{+0.08}_{-0.08}$ &  0.67 $\pm$ 0.09  & 1.3$^{+3.5}_{-0.8}$ & 4.4 &  1.38$^{+0.21}_{-0.22}$ &  -0.80$^{+0.10}_{-0.12}$  \\
NGC~4636 & 9  &  0.31$^{+0.09}_{-0.09}$ & 0.54 $\pm$ 0.11  &  0.4 $^{+0.3}_{-0.2}$  &   0.34 &  $0.30^{+0.24}_{-0.24}$ &   -1.52$^{+0.10}_{-0.13}$ \\
M~60     & 83 & 1.0$^{+0.1}_{-0.2}$   & 0.80 $\pm$ 0.02   & 0.16$^{+0.05}_{-0.04}$  & 11.0  &   $1.79^{+0.24}_{-0.24}$  &  -0.27  \\ 
\hline 
\end{tabular}
 
\medskip
Column description:  
    1) name;  2) Bondi radius in pc;  3) slope of
  the density profile;  4) temperature at the Bondi radius in keV; 
    5) number density at the Bondi radius in cm$^{-3}$;  6) Bondi
  accretion rate in $0.001 \times M_{\odot}$ yr$^{-1}$;  7)
  accretion power in $10^{43}$ erg s$^{-1}$;  8) jet power in  $10^{43}$ erg s$^{-1}$ 
  (when not reported a fixed error of 0.4 dex has been assumed, the other
  values are from \citealt{allen06}).
\label{tab2}
\end{table*}

\begin{table*}
\caption{Results of the statistical and correlation analysis.
}
\begin{tabular}{|c|c|c|c|c|c|c|c|c|}
\hline 
X &  Y   &  N &  r  & $\rho$ & $P_{\rho}$  &  a   &   b   &  rms \\
\hline
$P_B^V$ & $P_J^V$ & 10  & 0.94 & 0.79  & 0.018  & 1.17 $\pm$ 0.16 &
-2.13 $\pm$ 0.23  & 0.25  \\
$P_B$ & $P_J$ & 27 & 0.81 & 0.79  & $  10^{-4}$  & 1.05 $\pm$ 0.13 &
-1.96 $\pm$ 0.27  & 0.51  \\
$P_B^{V,\star}$ & $P_J^V$ & 10  & 0.93 &   0.79 & 0.03  & 0.91 $\pm$ 0.17 &
-1.38 $\pm$ 0.16  &  0.29 \\
$P_B^{\star}$ & $P_J$ & 27 & 0.62  & 0.64  & 0.002  & 1.27 $\pm$ 0.21 & -1.24
$\pm$ 0.27  & 0.69  \\
$M_{BH}^V$ & $P_J^V$ & 10  & 0.86 & 0.80  & 0.016  & 1.71 $\pm$ 0.36 &
-15.6 $\pm$ 3.1  & 0.37  \\
$M_{BH}$ & $P_J$ & 27 & 0.59 & 0.64  & $ 1.2 \times 10^{-3}$  & 2.20 $\pm$ 0.41 &
-19.5 $\pm$ 3.6  & 0.73  \\
$n^V_{1\, kpc}$ & $P^V_B$ & 10 & -  & 0.49  & 0.14 &  - & - & -  \\
$n_{1\, kpc}$ & $P_B$ & 27 &  0.42 & 0.39 & 0.05 & 1.67 $\pm$ 0.38 & 4.10
$\pm$ 0.52  & 0.80  \\

\hline 
\end{tabular} 

\medskip
Column description: 1) dependent variable, and 2) independent variable 
(superscripts $V$ indicate that only the Virgo sample is considered); 3) number
of objects; 4) linear correlation coefficient; 5) Spearman rank correlation; 
6) probability of obtaining the observed (or a
  higher) value of $\rho$ under the null hypothesis of no correlation between
  the two variables; 
7) slope, and 8) intercept of
the correlation (Y  = a X + b; the two upper limits on $P_J$ for NGC~4365 and
M~85 are not considered);  9) scatter of the correlation. 

$^{\star}$ The density $n_{\rm inn}$ measured at the innermost annulus (see text) is adopted to estimate $P_B$.
\label{tab3}
\end{table*}
\section{Jet and accretion powers}
In this section we estimate the jet's kinetic energies and the accretion
powers of the four CoreG with the lowest radio luminosity.  The main
assumptions and the details of the methods adopted to obtain $P_B$ and $P_J$
are outlined and discussed in BBC08 (and references therein).

\begin{figure*}
\centerline{
\psfig{figure=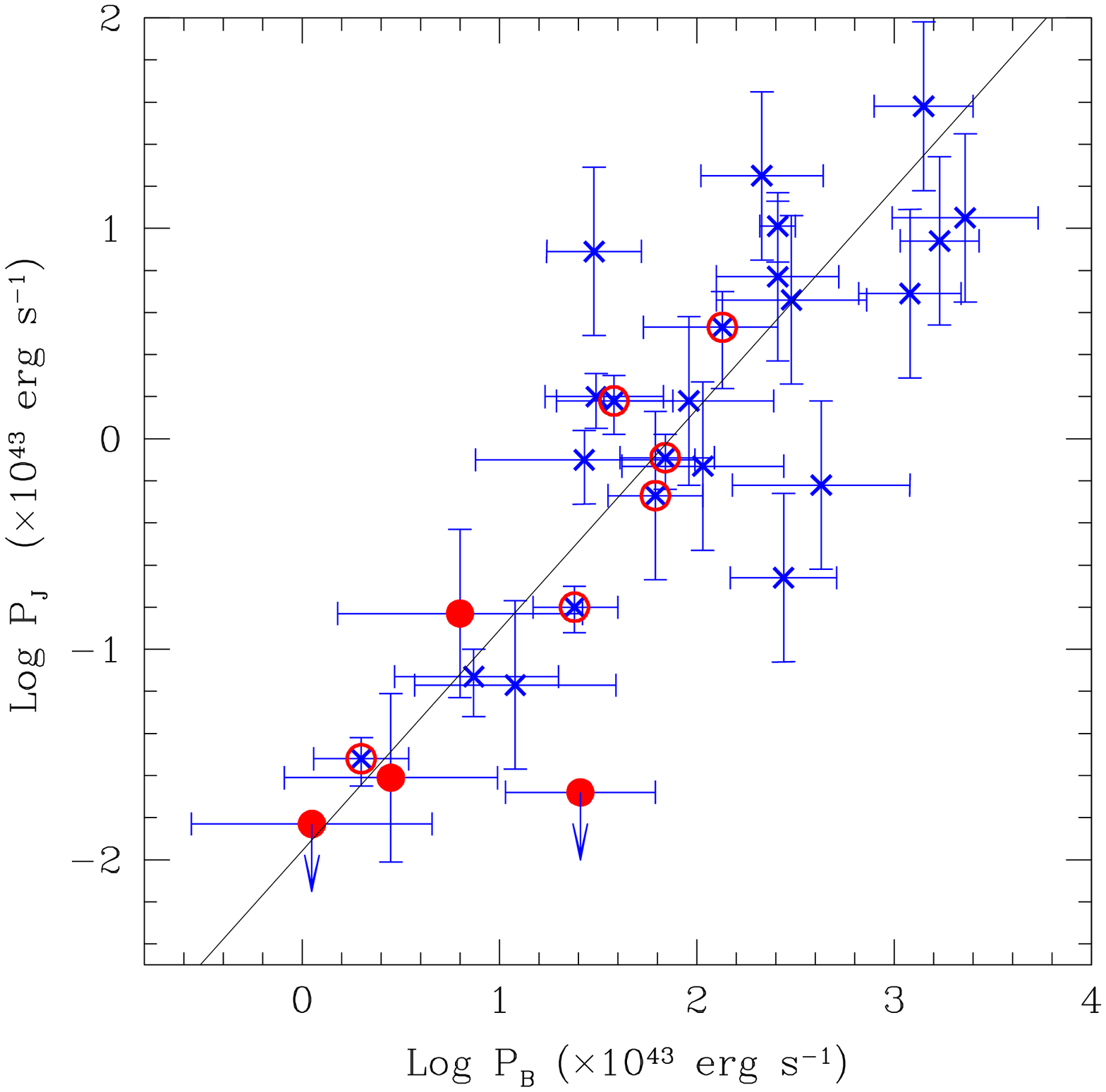,width=0.50\linewidth}
\psfig{figure=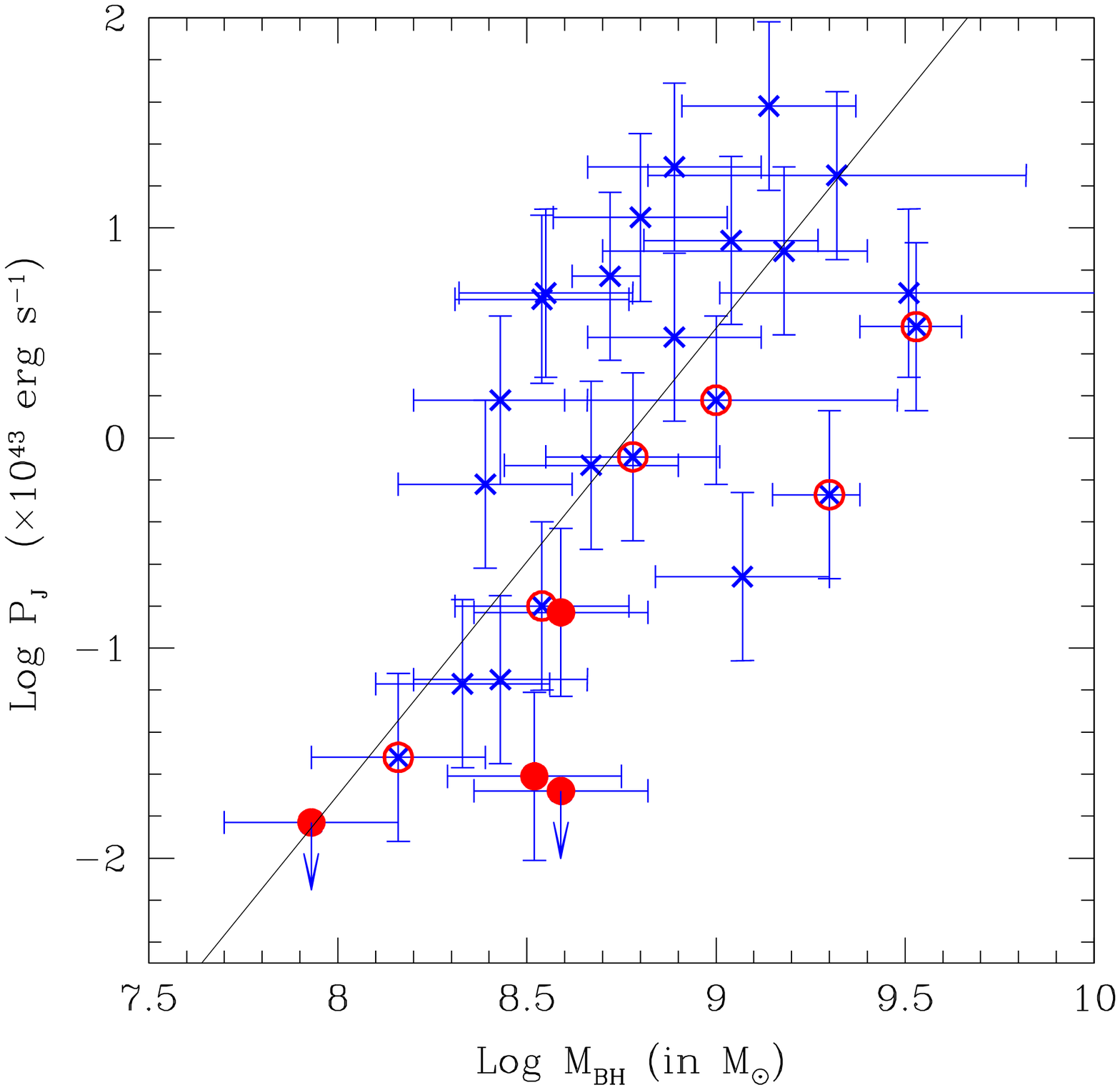,width=0.50\linewidth}
}
\caption{Jet power $P_j$ vs the accretion power $P_B$ ({\it left}) and the
  mass of the SMBH $M_{BH}$ ({\it right}) for the sample of 27 objects,
  consisting of the 23 objects from BBC08 and \citet{allen06} with the
  addition of the four new measurements. The 10 sources belonging to the Virgo
  cluster are marked with red circles, while full circles indicate the four
  objects analyzed here. The solid lines represent the best-fit
  regressions for the whole sample.}
\label{pjpb}
\end{figure*}

\subsection{Jet power}
As already mentioned in the introduction, the jet kinetic power $P_J$ can be
estimated from the work done by the jets to inflate the ISM cavities observed
in the X-ray images. This method can be used only in objects with high X-ray
data quality and, obviously, when such bubbles are
visible. \citet{heinz07} showed that the radio core luminosity, $L_{\rm core}$,
can also be used to estimate $P_J$. Using a sample of radio-galaxies for which
$P_J$ could be measured from the presence of cavities, they calibrated a
relation in the form $P_J \propto L_{\rm core}^{12/17}$ that enables one to
evaluate $P_J$ from a simple measurement of core radio power $L_{\rm core}$,
with a typical uncertainty (based on the rms of the relation) of $\sim$ 0.4
dex.

The four galaxies under scrutiny have rather different radio properties with
respect to those considered by in previous studies. Indeed, while the galaxies
considered in \citet{allen06} and BBC08 have well-developed double radio
structures, only NGC~4526 has an extended morphology indicative of a jet
origin. The radio emission of M~86 is instead unresolved, while it is
undetected in NGC~4365 and M~85. The very presence of jets in these last three
sources cannot be taken for granted. We might refer more precisely to our
analysis of the connection between accretion and jet power as a study of the
link between accretion and the AGN radio emission.  Nonetheless, because
$L_{\rm core}$ and $P_{\rm jet}$ differ only for the scaling relation reported
above, we prefer to maintain the jet power as a reference quantity, to
facilitate comparison with previous works.

This assumes that the $P_j$ - $L_{\rm core}$ relation holds for the sources
discussed here, with a core power at least 10 times fainter than those
considered by \citet{heinz07}, i.e. that their jets properties are similar to
those of brighter radio-galaxies. In particular, there is the possibility that
low-power jets are substantially slower, even non relativistic. Implicitly
adopting the average de-boosting factor for radio cores of more powerful
sources for our sources, $\delta^2 = 4 \times \Gamma^{-2} (2 - \beta)^{-2}$,
might cause an overestimate of the jet kinetic power of up to a factor of
$\sim 3.6$ (adopting $\Gamma = 5$ from \citealt{giovannini01}). Although this
is not a negligible effect, it does not significantly affect our main
conclusions.

Because the estimates of $P_J$ based on cavities can be considered as primary
calibrators, while those derived from $L_{\rm core}$ are secondary estimators,
we preserved these more direct measurements for the nine objects of Allen et
al. for the analysis that will follow in Sect. \ref{discussion}.

\subsection{Accretion power}
The accretion power $P_B = \eta \dot M_B c^2$ is deduced assuming a steady,
spherically symmetric model for accretion onto a gravitational body
\citep{bondi52}.  For the sake of simplicity we have assumed an efficiency $\eta =
1$; the accretion rate, $\dot M_B$, is defined at the Bondi radius $r_B = 2 G
M_{BH}/c^2_s$ ($c_s$ is the sound speed, $M_{BH}$ the mass of the SMBH) as

\begin{equation}
\dot M_B \propto c_s \rho_B r^2_B  \propto  n_B M^2_{BH} T_B^{-3/2}
  \label{eq:MBdot},
\end{equation}
where $\rho_B$, $n_B$, and $T_B$ are the mass and number densities, and the
temperature at $r_B$, respectively.

The mass of the SMBH has been estimated from the stellar velocity
dispersion, available for all sources from the HyperLeda database, using the
\citet{tremaine02} relation.
In three CoreG of the Virgo sample, the mass of the SMBH has been 
measured directly and we adopted the value reported by
\citet{marconi03}.

The temperature and density were deduced from the properties of the hot
ISM through Chandra X-ray observations. The data analysis and the results of
these observations are presented in detail in the appendix, where we derive
the deprojected profiles of temperature and density across the coronae for
the four VCC sources. The temperature profiles across the coronal regions are
essentially constant, with typical values of $T \approx 0.4 - 0.7$ keV. For
the calculation of the Bondi accretion rate we assumed that $T_B$ is equal to
the temperature at the innermost shell. The value of $n_B$ was
extrapolated down to $r_B$ assuming a relation $n(r) \propto r^{-\alpha}$,
with $\alpha$ deduced from a power-law fit to the density profile of the ISM
(see the appendix for details). 

The values of these quantities, with the derived accretion and jet powers, are
reported in Table \ref{tab2}, which also includes the data for the other Virgo
objects from BBC08.\footnote{A few misprints in their Table 2 have been
  corrected.} We remark that for our four VCC objects the quality of the data is
such that we were able to derive an estimate of $\dot M_B$ with a typical
uncertainty of a factor of 3. Besides these purely statistical uncertainties,
systematic effects can also play a role. For example we
will discuss in the next Section the effects of the assumption of the power-law dependence of the
density on radius.

\section{Discussion}
\label{discussion}

\subsection{Accretion versus jet power}
\label{pjpbsection}
In the left panel of Fig. \ref{pjpb} we compare the jet and the accretion
power, i.e. $P_J$ and $P_B$. The four radio-faint VCC/CoreG are all located in
the bottom left corner of this diagram. This indicates that their low radio
luminosity is associated with a low level of accretion and with values
consistent with the correlation between $P_J$ and $P_B$ derived for larger
powers. The results of the statistical analysis are reported in Table 3: 
the parameters of the linear correlations are deduced from  the bisectrix of
the fits interchanging the two related quantities (excluding upper limits) 
as independent and dependent variables \citep{isobe90}.

The inclusion of these four objects only marginally alters the
statistical parameters describing the $P_J$ vs. $P_B$ relation with respect to
the results of BBC08 (see our Table \ref{tab3} in comparison with their
Table 3), yielding
$${\rm log} ~P_{J,43} = (1.05\pm0.13) ~{\rm log} P_{B,43} - (1.96 \pm0.27),$$
where both powers are given in units of $10^{43}$ erg s$^{-1}$.

Focusing on the sub-sample of 10 Virgo galaxies, $P_B$ and $P_J$ are also well
correlated, with Spearman rank correlation $\rho = 0.79$, corresponding
  to a probability $P^V_{\rho} = 0.018$ of obtaining the observed (or a
  larger) value of $\rho$ under the null hypothesis of no correlation.  Slope
and intercept agree within the errors with those found for the full
sample. Because these sources represent the majority of the low power sources,
this indicates that the results of BBC08 also hold for very low accretion
rates and jet powers, and in particular that the conversion efficiency from
accretion to radio or jet power does not significantly vary over $\sim$ 4
orders of magnitude.

We also tested how critical our assumption is that the density $n_B$ can be
extrapolated using a relation in the form $n \propto r^{-\alpha}$ down to the
Bondi radius. Indeed, even though the density profiles in our sources appear
to monotonically increase toward the center (with the exception of M~87, see
\citealt{allen06}) in the inner regions which are not probed by the Chandra
data, they might flatten (as pointed out by \citealt{hardcastle07}). We then
explored the effects of adopting as $n_B$ the value of the density at the
innermost annulus of the profile ($n_{\rm inn}$), which can be considered as a
strict lower limit to the actual density at $r_B$. Despite the very large
range in the value of the innermost radius $r_{\rm inn}$ compared to the Bondi
radius, renormalizing the value of $P_B$ with $n_B = n_{\rm inn}$, the
correlation $P_J - P_B$ holds (with $P^V_{\rho}$ and $P_{\rho}$ = 0.03 and
0.002, respectively; see Table \ref{tab3}). The quasi linear dependence is
maintained, although with a slightly higher scatter, and with an intercept
(and thus a conversion efficiency) increased by a factor of $\sim$ 5.

Following BBC08, we looked for possible relationships between the jet
power and the single quantities defining $P_B$ through Eq. (1). No correlation
was found between $P_J$ and the temperature at the Bondi radius.  A
correlation $P_J - n_B$ emerges only when the whole sample is considered
($P_{\rho} = 0.03$) but with a large scatter (rms = $0.81$), an indication
that the highest jet powers are in general found in the densest (at their Bondi
radius) coronae. A stronger relationship of $P_J$ with $M_{BH}$ is conversely
present as shown by the right panel of Fig. \ref{pjpb} and in Table
\ref{tab3}. The slope of the fit ($a \approx 2.2\pm0.4$ for the full sample
and $a \approx 1.7\pm0.4$ for the Virgo sub-sample) agrees with that
predicted by the dependence of the accretion rate from $M^2_{BH}$. We must
remark however that the scatter of the correlation $P_J$ - $M_{BH}$ is larger
than for the $P_J$ - $P_B$ relation. From these arguments it is clear that the
mass of the SMBH is the dominant parameter in driving the accretion process
and powering the jet, but that the density at the Bondi radius plays an
important role, reducing the scatter of the correlation.

\begin{figure}
\centerline{
\psfig{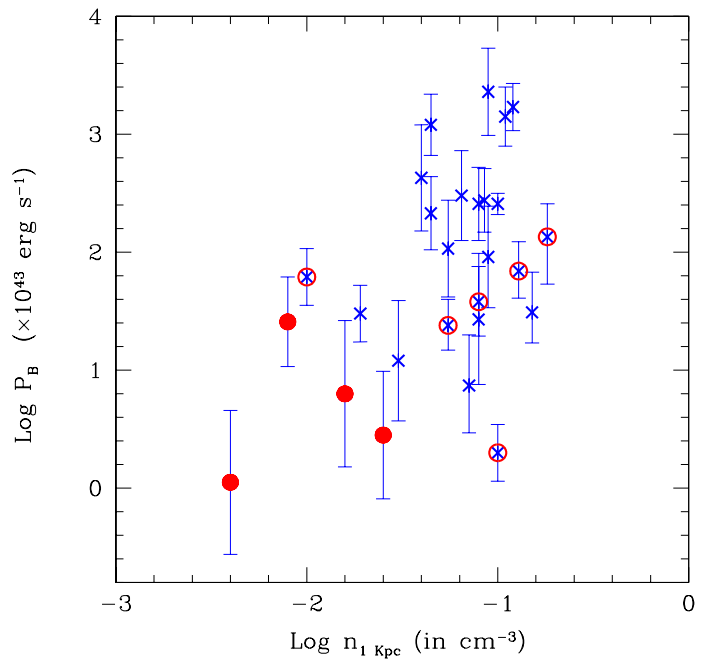}}
\caption{$P_B$ vs $n_{1\, kpc}$, the density of the ISM at 1 kpc. Symbols as
  in Fig. \ref{pjpb}, i.e. red circles represent the 10 sources belonging to
  the Virgo cluster, with full circles indicating the four new measurements.}
\label{pbalfa} 
\end{figure}

\subsection{$P_B$ and large scale properties of the ISM}
\citet{balmaverde08alias} found that the different levels of nuclear activity
are related to global differences in the structure of the galactic hot
coronae. Indeed, a relation links the jet power with the corona X-ray surface
brightness, albeit with a large scatter. This suggests that a substantial
variation in the jet power must be accompanied by a global change in its ISM
properties. As the temperature is quite uniform across the ISM, the density in
the external region is the most relevant parameter that could be related to
the accretion process occurring in the core.  We here re-examine this issue,
but instead of considering the surface brightness of the corona, we measured
its de-projected density at a fixed radius of 1 kpc, $n_{\rm 1 kpc}$, $\sim$ 1
-- 2 orders of magnitude larger than the Bondi radius, interpolating the
neighboring points in the density profile. In Fig. \ref{pbalfa} we compare
$n_{\rm 1 kpc}$ with $P_B$.

For the complete sample of 27 galaxies, the correlation analysis between $P_B$
and $n_{\rm 1 kpc}$ provides $P_{\rho} = 0.05$ and $a \approx 1.7 \pm
0.4$.\footnote{No convincing correlation is found though considering only the
  10 Virgo galaxies, see Table 3.} The residuals have a rms of 0.8 dex. A
similar relation is present also between $P_J$ and $n_{\rm 1 kpc}$.  The
inclusion of the four low-power VCC galaxies is crucial to unveil the
connection between these quantities, because it extends the range covered by
both variables, providing the necessary leverage to the data. We thus confirm
that the level of AGN activity is connected with the overall properties of the
hot corona.

The rather loose relation between accretion rate and coronal density is not
surprising. Indeed, assuming typical sound speeds of $\sim 500$ km s$^{-1}$,
the outer (1 kpc) and inner regions ($r_B \sim 10 - 100$ pc) of the ISM are
causally connected on dynamical timescales on the order of a few $10^6$
years. The large scatter observed can be related to events occurring inside
the ISM on shorter timescales. These could be of external origin,
e.g. transient changes due to mergers or interactions with the ISM, but
the accretion flow can also be affected `from inside', e.g. due to a variable
release of jet kinetic energy into the corona.

\begin{figure*}
\centerline{
\psfig{figure=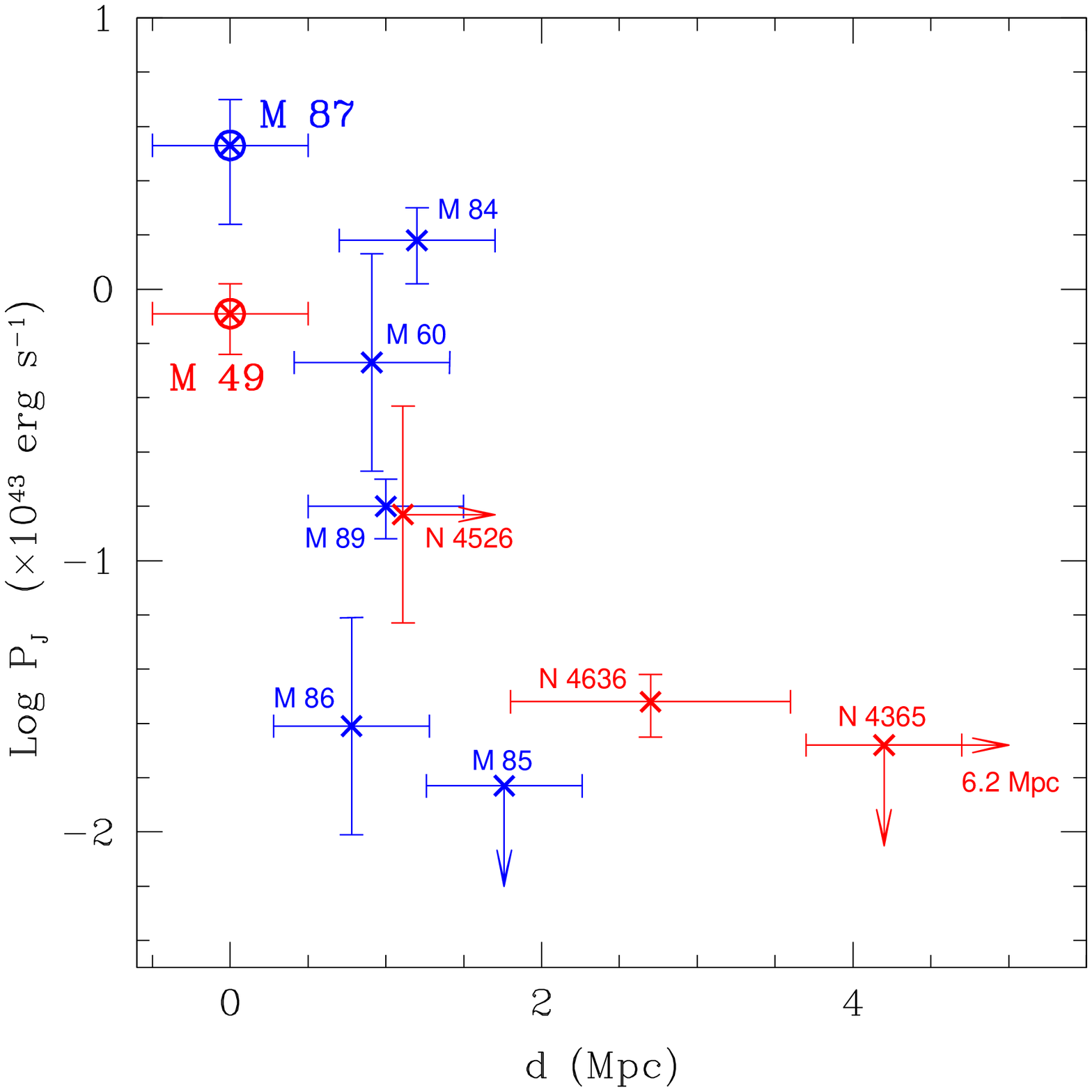,width=0.50\linewidth}
\psfig{figure=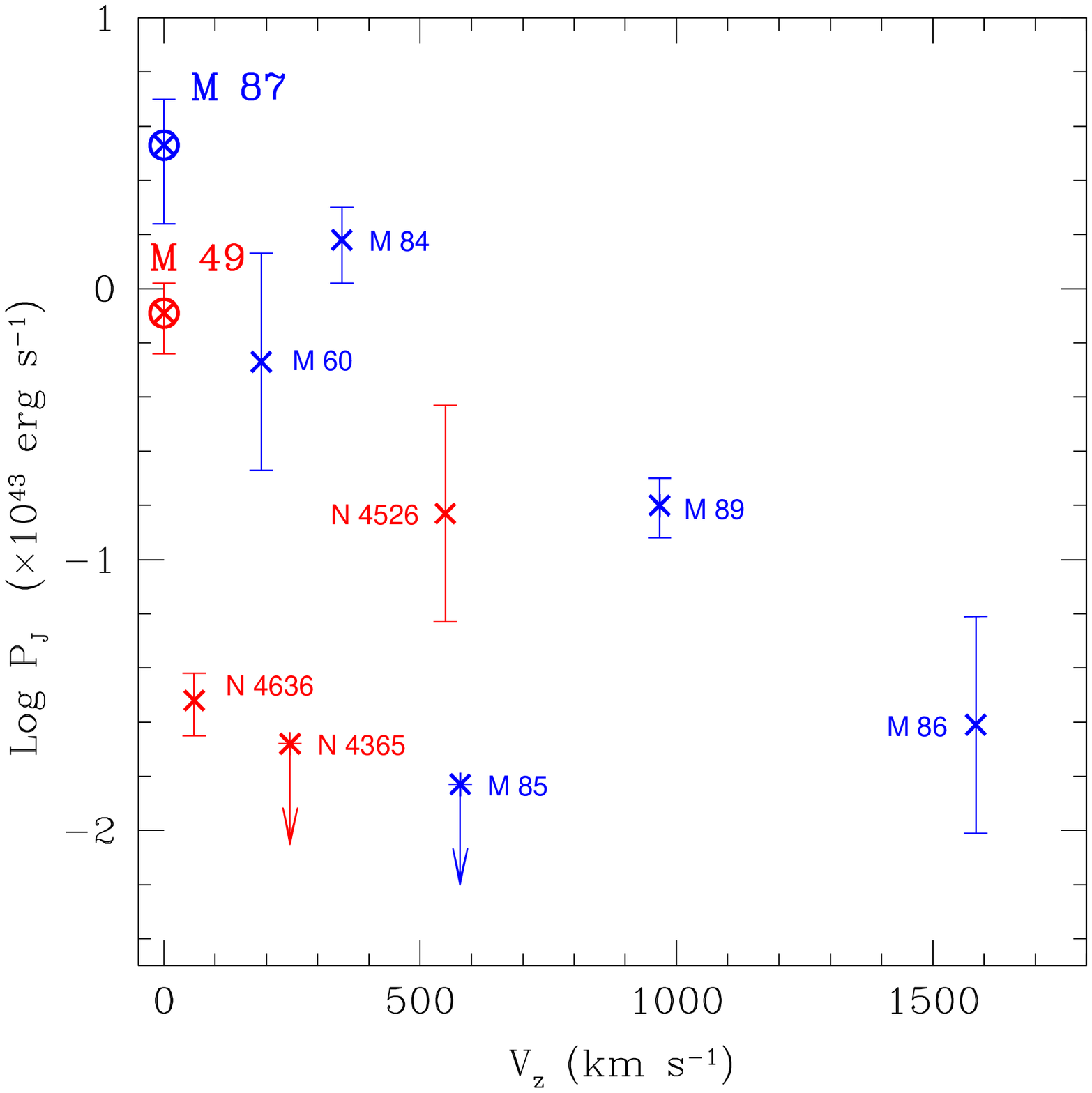,width=0.50\linewidth}
}
\caption{$P_J$ vs $d$, the relative three dimensional distances (left panel),
  and $V_z$, the radial velocities (right panel), of the VCC objects from the
  dominant components in Virgo A (blue) and Virgo B (red). Because the
  distance of NGC~4526 from the Galaxy cannot be measured using the brightness
  fluctuation method (due to a prominent dust lane) we only have a lower limit
  for $d$. NGC~4365 is instead indicated as an arrow with its large distance
  (6.2 Mpc) reported explicitely for easier visualization.}
\label{fig4} 
\end{figure*}

\subsection{Accretion and the structure of the Virgo Cluster}

We now look for possible relationships between the jet power 
(strictly connected to the
nuclear activity) of the Virgo CoreG and their location within the cluster,
and in particular with respect to the dominant galaxies of Virgo A and B (see
Table \ref{tab1}). In Fig. \ref{fig4} we compare $P_J$ with (left panel)
the three dimensional distance of each galaxy with respect to its relative
sub-cluster dominant component and (right panel) also in terms of difference
of recession velocity.

A suggestive trend between distance from the dominant galaxy of each
subcluster and jet power emerges; galaxies located closer to the center of
each sub-cluster have a larger jet power than those at larger distances.  Due
to the large uncertainties and to the low number of objects this does not
anyway correspond to a statistically significant correlation between $P_J$ and
$d$.  An analogous trend appears when considering the absolute value of the
relative radial velocities, in the sense that high jet powers are not found in
objects with the higher velocities. Clearly this analysis is limited because
we are only able to measure one of the velocity components.

The clearer picture is derived by looking at distance and velocity
simultaneously. The case of M~86 is particularly instructive. It is located at
$\sim$ 1 Mpc from M~87, at a distance similar to that of M~84, M~89, and M~60.
With respect to these sources, M~86 has a jet power $\sim 6 - 60$ times lower
(and a similarly lower accretion rate). This could be due to different
dynamical effects in these galaxies. Indeed, M~86 has the highest radial
relative velocity with respect to M~87 ($V_z \approx 1600$ km s$^{-1}$, being
actually blue-shifted with respect to the galaxy) and its Chandra image shows
a long tail in the NW direction \citep{randall08}, a clear evidence of ram
pressure stripping. The coronal gas content is progressively depleted at each
passage in the central regions of the cluster, possibly causing the lower
level of activity.

From the available data it is difficult to separate these environmental
effects from a possible mass segregation that would cause the more massive
galaxies to be preferably found closer to the cluster's center. These would
have correspondingly higher SMBH masses that due to the important role played
by $M_{BH}$ might be associated with objects with higher accretion and higher
jet power.

\section{Summary and conclusions}
We have extended previous studies on the connection between accretion of hot
gas and jet power in radio-loud AGN with a sample of four objects belonging to
the Virgo cluster, characterized by extremely low radio luminosity.  Two of
them are actually undetected in 8.4 GHz VLA observations, corresponding to a
limit to their radio luminosity of $L_{\rm core} \lesssim 3 \times 10^{25}$
erg s$^{-1}$. These four galaxies are particularly interesting because they
have similar properties to those typical of the hosts of radio-loud AGN from
the point of view of their brightness profiles (they are giant early type
galaxies with a flat nuclear stellar core) and are associated with black holes
of large mass ($M_{\rm BH} \sim 10^8-10^{8.5}$M$_{\sun}$). Nonetheless,
these objects are $\gtrsim 30,000$ times fainter than the radio core of M~87
(the brightest VCC galaxy) and $\gtrsim 200,000$ times fainter in total radio
power. Among the various alternatives that can be proposed we here explore the
possibility that this is due to an extremely low accretion rate.

We used Chandra observations of these sources to estimate the power available
from accretion of hot gas, $\dot M_B$, in a spherical approximation. This was
obtained by de-projecting the temperature and density profiles of each source.
Because the diffuse emission in these targets is on average much fainter than
in those considered in previous studies, we considered and removed the
contamination from stellar sources to isolate the genuine ISM emission. The
quality of the data is such that we could estimate $\dot M_B$ with a typical
uncertainty of a factor of 3.

Furthermore, we now have a complete sample formed by all 10 core-galaxies
located in the Virgo cluster. It has then been possible to verify whether the
connection between jet and accretion rates holds also in this environment and
whether galaxies show different properties, depending on their positions
inside the cluster.

Our results can be summarized as follows.

\noindent {\bf 1.} 
The estimated accretion rates of the four radio-faint VCC galaxies are
among the lowest of the sample considered. This indicates that their low radio
luminosity is most likely due to a correspondingly low level of accretion.

\noindent {\bf 2.} The correlation between the accretion and the jet powers
spans over 4 decades down to very low values of $P_B$ and $P_J$, with an
approximately constant conversion rate of $P_J \sim 0.01 \times P_B$. This
even holds relaxing our assumption that the density profile can be
extrapolated with a power-law to estimate its value at the Bondi radius. The
mass of the SMBH is the main driving parameter of the correlation.
 
\noindent {\bf 3.} The correlation between $P_B$ and $P_J$ is also found for
the subsample of the 10 Virgo sources, with parameters similar to those
describing the full sample.

\noindent {\bf 4.} A rather loose relation links the accretion rate and
coronal density at 1 kpc, suggesting that a substantial variation in the jet
power must be accompanied by a global change in its ISM properties.

\noindent {\bf 5.} A suggestive trend (that however does not correspond
  to a statistically significant correlation) is present between the location
  (and relative velocity) of a source with respect to the cluster's center and
  its jet power. In one case, this is clearly due to ram pressure stripping
of hot coronal gas from a galaxy moving at high speed with respect to the
cluster.

The main findings of this work are additional support for a linear relation
between the power available from accretion of hot gas and the jets kinetic
power and that the efficiency of this process is apparently independent of the
level of luminosity, of the black hole mass, and of the environment. Although
we are still far from understanding the dynamics of the accreting gas close to
the SMBH, these results represent robust constraints on the processes of jets
formation.

With the current instrumentation it is unlikely that a much larger range of
AGN luminosity can be probed, considering (at the low end) the depth of the
available radio observations and (at the high end) the limit imposed by the
spatial resolution of the X-ray images in more distant and powerful AGN. 

It would instead be important to detect the two galaxies still elusive in our
VLA survey in deeper radio images (currently within reach thanks to the
Extended VLA) to confirm (or disprove) that they are genuine active
galaxies, even though of very low luminosity. They can be very useful to probe
the properties of accretion and jet production at extreme low powers, in
particular NGC~4365 which is already a marginal outlier in the $P_J - P_B$
diagram.

In the objects considered until now the accretion process is highly
radiatively inefficient and the electromagnetic output from the nucleus is
negligible with respect to what emerges as kinetic power (BBC08). The
situation is rather different for radio-quiet AGN, which are brighter by a
factor $\sim 100$ than radio-loud AGN in the X-ray band at equal radio power
(e.g. \citealt{panessa07}). It would be of great interest to explore whether,
in analogy to the $P_J - P_B$ relation in radio-loud sources, a correlation
between the accretion power and radiative output (e.g. a link between $P_B$
and $L_X$) exists in radio-quiet objects. An initial attempt in this direction
was made by \citet{soria06a,soria06b} analyzing a sample of eight galaxies
selected for their low X-ray luminosity. No correlation was found, but these
objects span only a factor of 10 in $P_B$ and $L_X$.  An extension to
radio-quiet sources covering a much larger range of X-ray luminosity is
clearly needed to firmly test this point.

\begin{acknowledgements} The authors thank the anonymous referee and Barbara
  Balmaverde for helpful suggestions about the data analysis and acknowledge
  partial financial support from ASI grant I/023/050. This publication makes
  use of data products from the Two Micron All Sky Survey (which is a joint
  project of the University of Massachusetts and the Infrared Processing and
  Analysis Center/California Institute of Technology, funded by the National
  Aeronautics and Space Administration and the National Science Foundation)
  and of the NASA/IPAC Extragalactic Database (NED) (which is operated by the
  Jet Propulsion Laboratory, California Institute of Technology, under
  contract with the National Aeronautics and Space Administration), and of the
  HyperLeda database (http://leda.univ-lyon1.fr).

\end{acknowledgements}

\appendix

\section{X-Ray data and ISM properties of the four radio-faint VCC galaxies}

The Chandra Archive information for the four VCC galaxies analyzed here are
reported in Table \ref{chandradata}. For the X-ray data reduction and analysis
we followed the same procedures and software packages as in BBC08. To
  account for the possible contamination from the active nucleus, we excluded
  the innermost region from the analysis, adopting a minimum radius of
  1\farcs5. This is rather conservative, because a faint X-ray central point
  source is seen only in NGC~4365 \citep{sivakoff03}.

  Since the diffuse emission in these targets is on average much fainter than
  in those discussed in BBC08, the contamination from stellar sources must be
  taken into account. These sources mainly belong to two classes: low-mass
  X-ray binaries (LMXB) and stars of old populations with hot coronae, mainly
  cataclysmic variables (CVs) and active binaries (ABs).

\begin{table}
 \caption{Chandra observations log.}
\begin{tabular}{|l|l|c|c|c|}
\hline
Source   & Alt. name& ObsId& Date        & Exp.(ks) \\
\hline  
VCC~~731 & NGC~4365 & 5921 & 04/28/2005  & 40.0 \\
VCC~~798 & M~85     & 2016 & 05/29/2001  & 40.3 \\
VCC~~881 & M~86     & 0963 & 04/07/2000  & 14.9 \\
VCC~1535 & NGC~4526 & 3925 & 11/14/2003  & 44.1 \\
\hline
\end{tabular} 
\label{chandradata}
\end{table}

The brightest LMXB appear in the X-ray images as discrete sources superimposed
to the diffuse coronal emission: they have been localized and removed from the
event files through the {\it wavedetect} command. However, less luminous,
unresolved LMXB can still have a non negligible contribution to the extended
luminosity. As their emission is well described by a power-law model, they can
be separated from the hot gas thermal emission through a spectral
decomposition. 

\begin{figure}
\centerline{
\psfig{figure=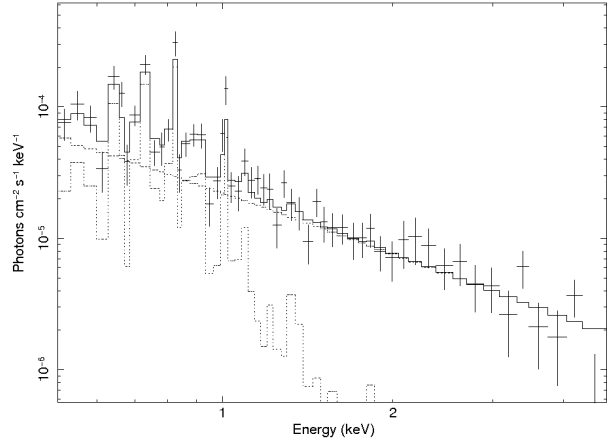,width=1.0\linewidth}}
\caption{Spectral fit to the spectrum of NGC~4365 adopting a
  thermal (Mekal) plus a continuum power-law composite model, representative
  of the ISM diffuse and LMXB emission, respectively.}
\label{figa1a}
\end{figure}

For this purpose we performed a two-components spectral fit, {\it Mekal +
  Pow Law}, adopting the galactic values for the column densities $N_H$. For
the thermal model the metalicity is fixed to half the solar abundances, while
for the power-law the spectral index has been fixed to the value found from
the discrete sources for each galaxy. The resulting power-law photon indices
are in the range $\Gamma \approx 1.3 - 1.6$, in good agreement with the
measurements of \citet{irwin03}.  We repeated this procedure considering four
annuli for each galaxy.  In three objects the integrated contribution of the
unresolved LMXB is at a similar level of the thermal emission from the ISM,
while it is negligible in NGC~4526. As an example we show in Fig. \ref{figa1a}
the results obtained for NGC~4365.

\begin{figure}
\centerline{
\psfig{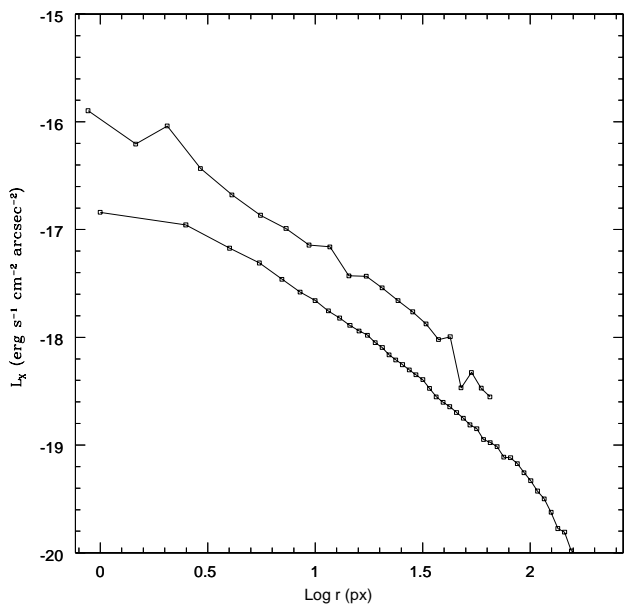}}
\caption{X-ray brightness profile of NGC~4365, as obtained from the Chandra
  observations (upper curve). The lower curve is an estimate of the
  contribution of X-ray emission from old active stars calculated scaling the
  K band surface brightness profile by using Eq. A.1.}
\label{figa1b}
\end{figure}

\begin{figure*}
\centerline{
\psfig{figure=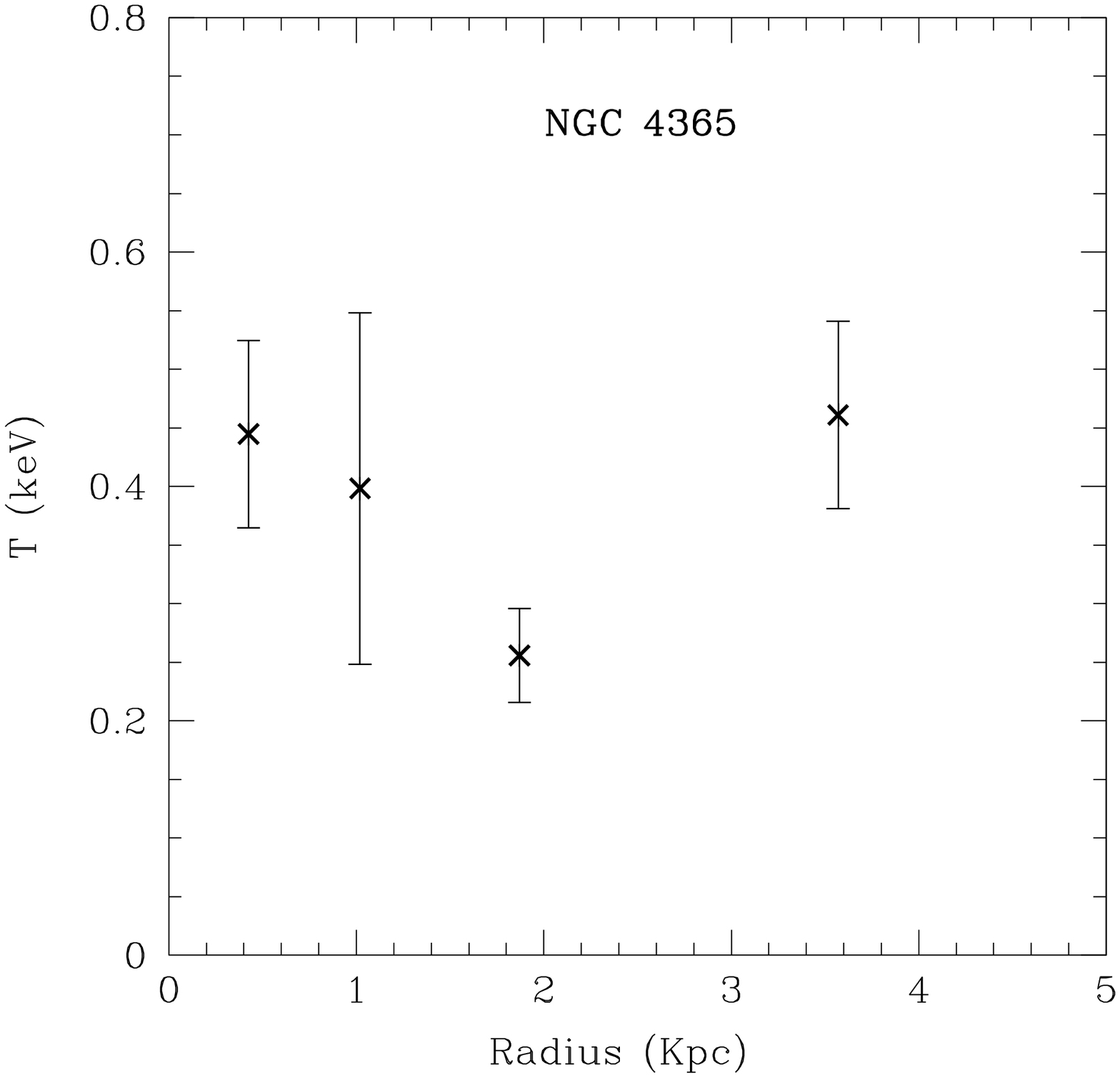,width=0.33\linewidth}
\psfig{figure=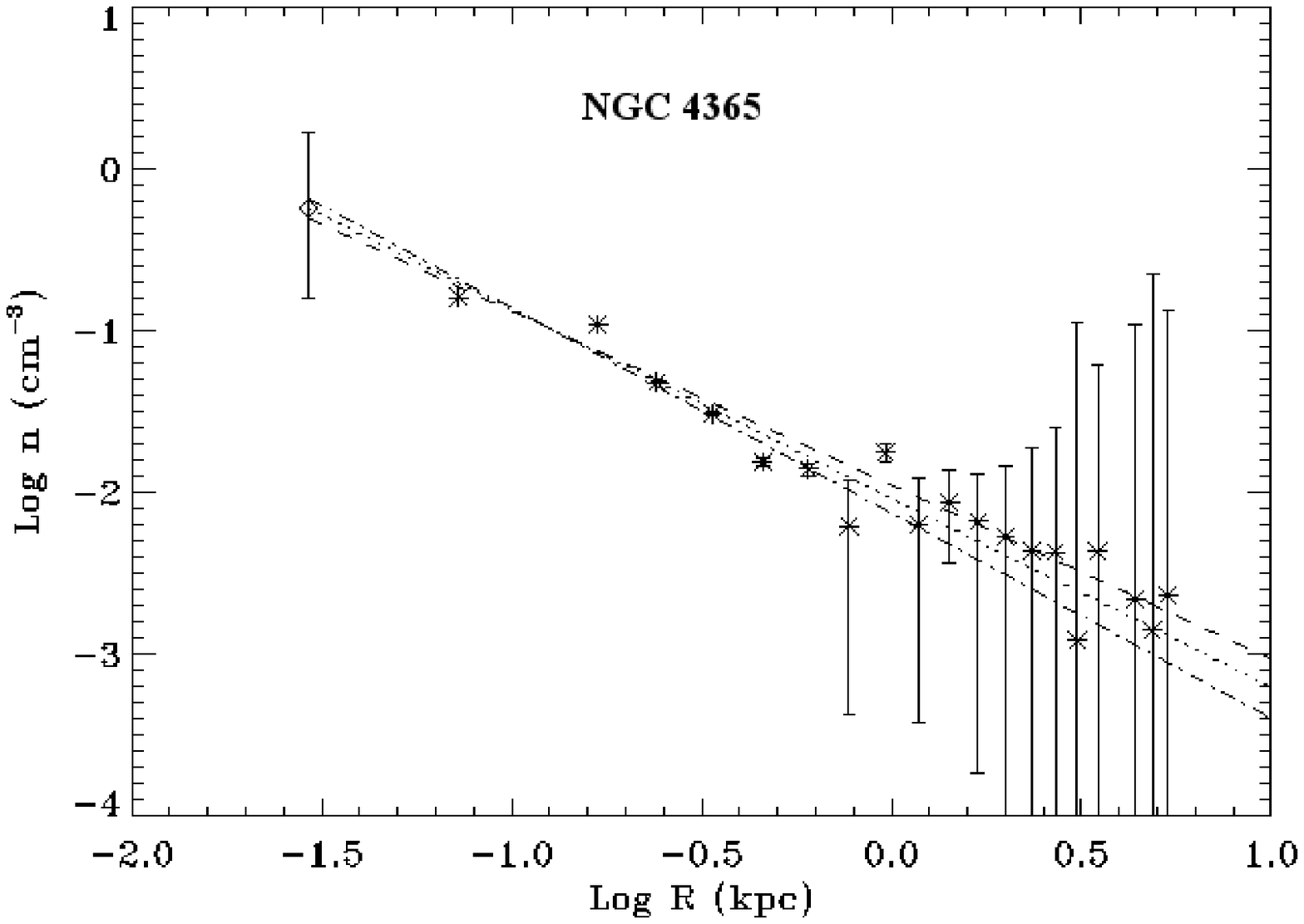,width=0.42\linewidth}}
\centerline{
\psfig{figure=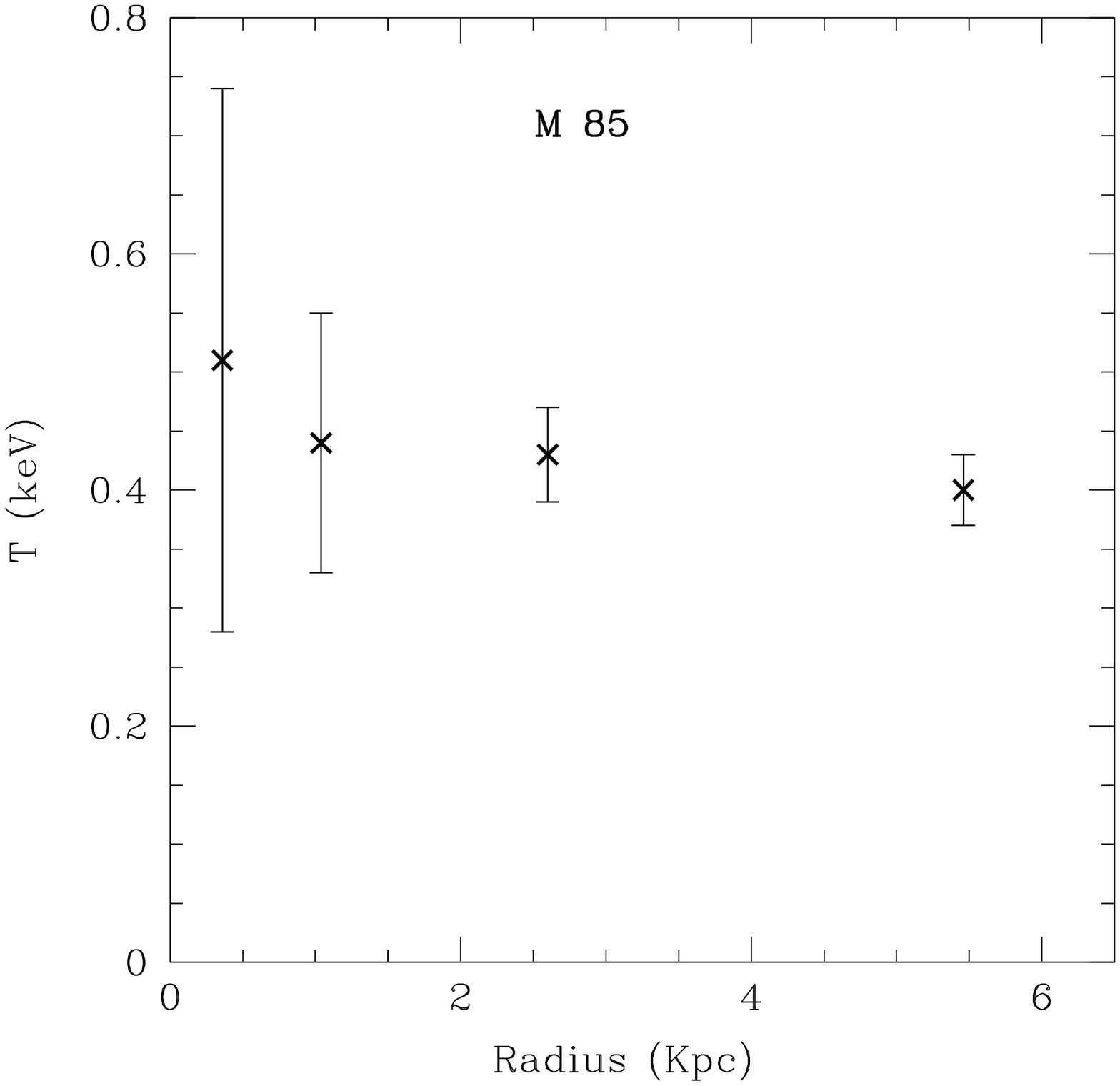,width=0.33\linewidth}
\psfig{figure=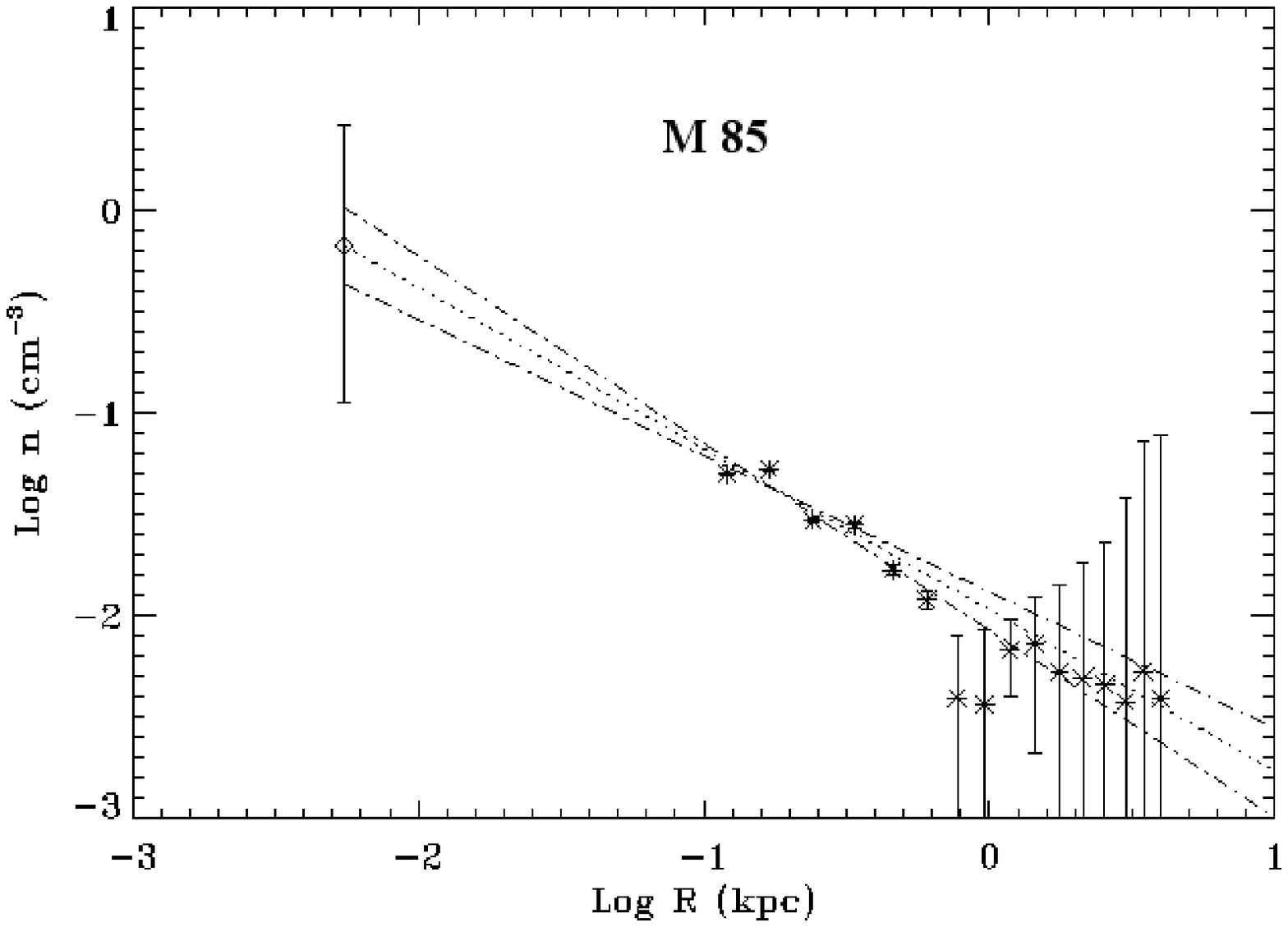,width=0.41\linewidth}}
\centerline{
\psfig{figure=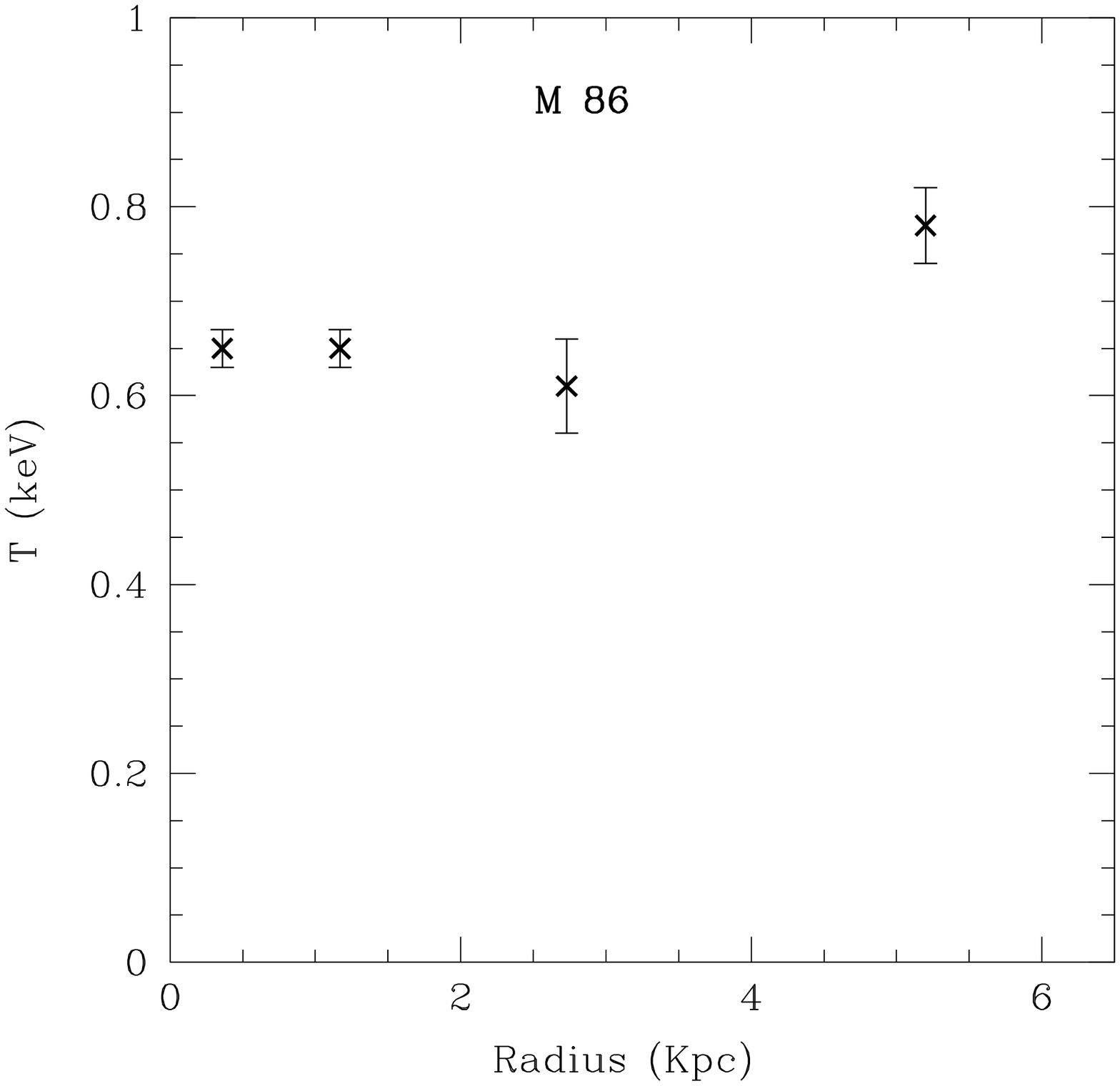,width=0.33\linewidth}
\psfig{figure=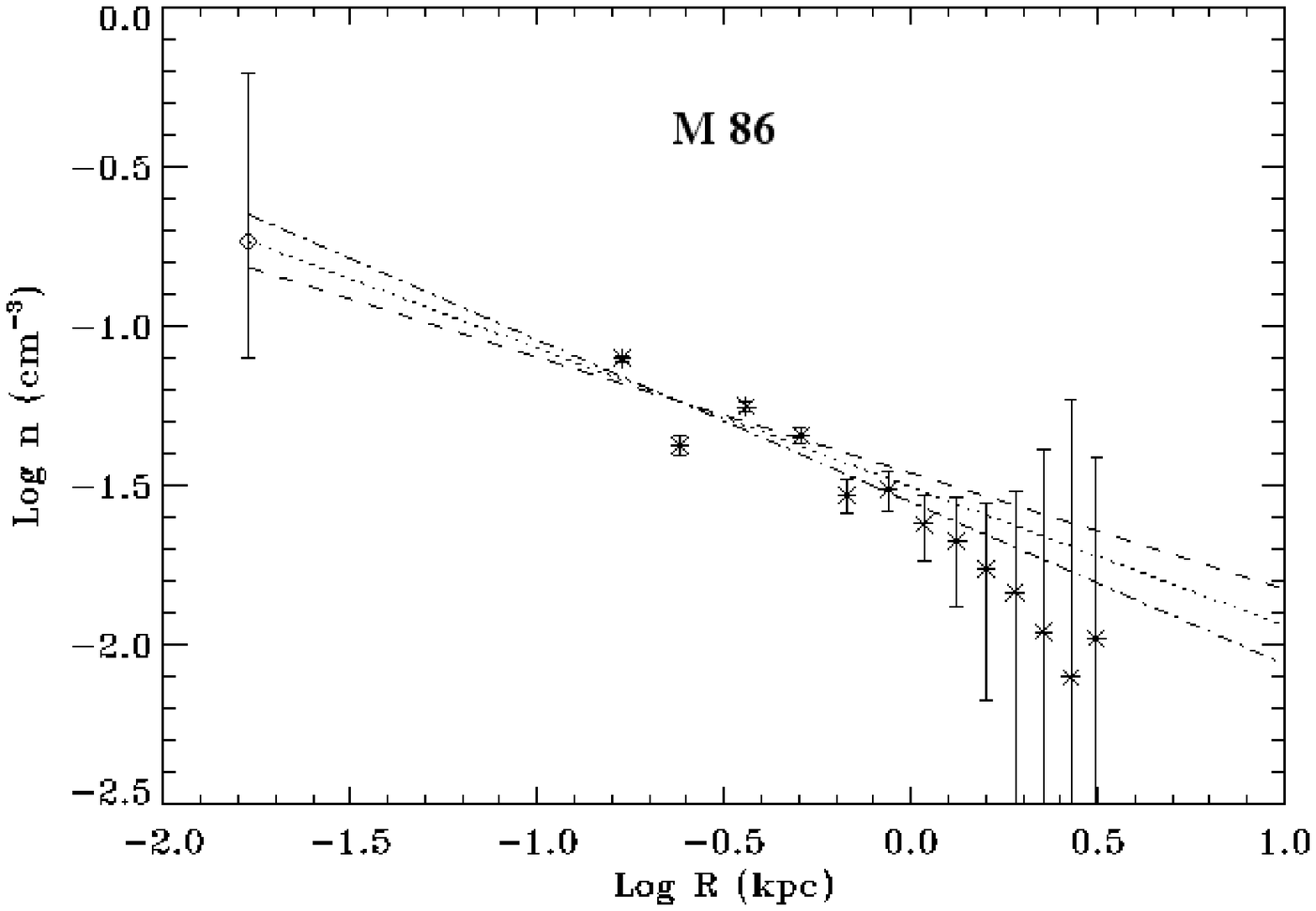,width=0.43\linewidth}}
\centerline{
\psfig{figure=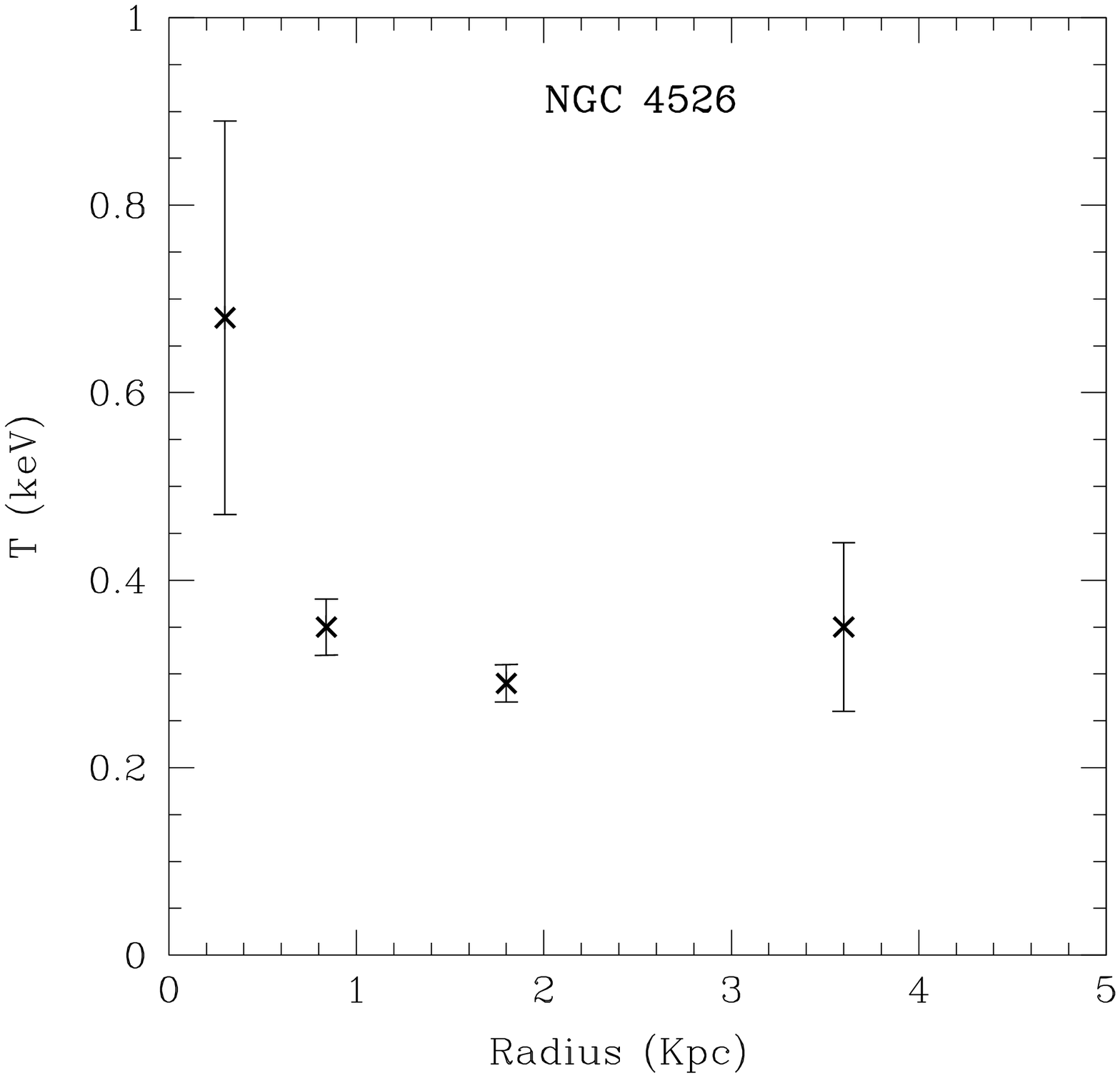,width=0.33\linewidth}
\psfig{figure=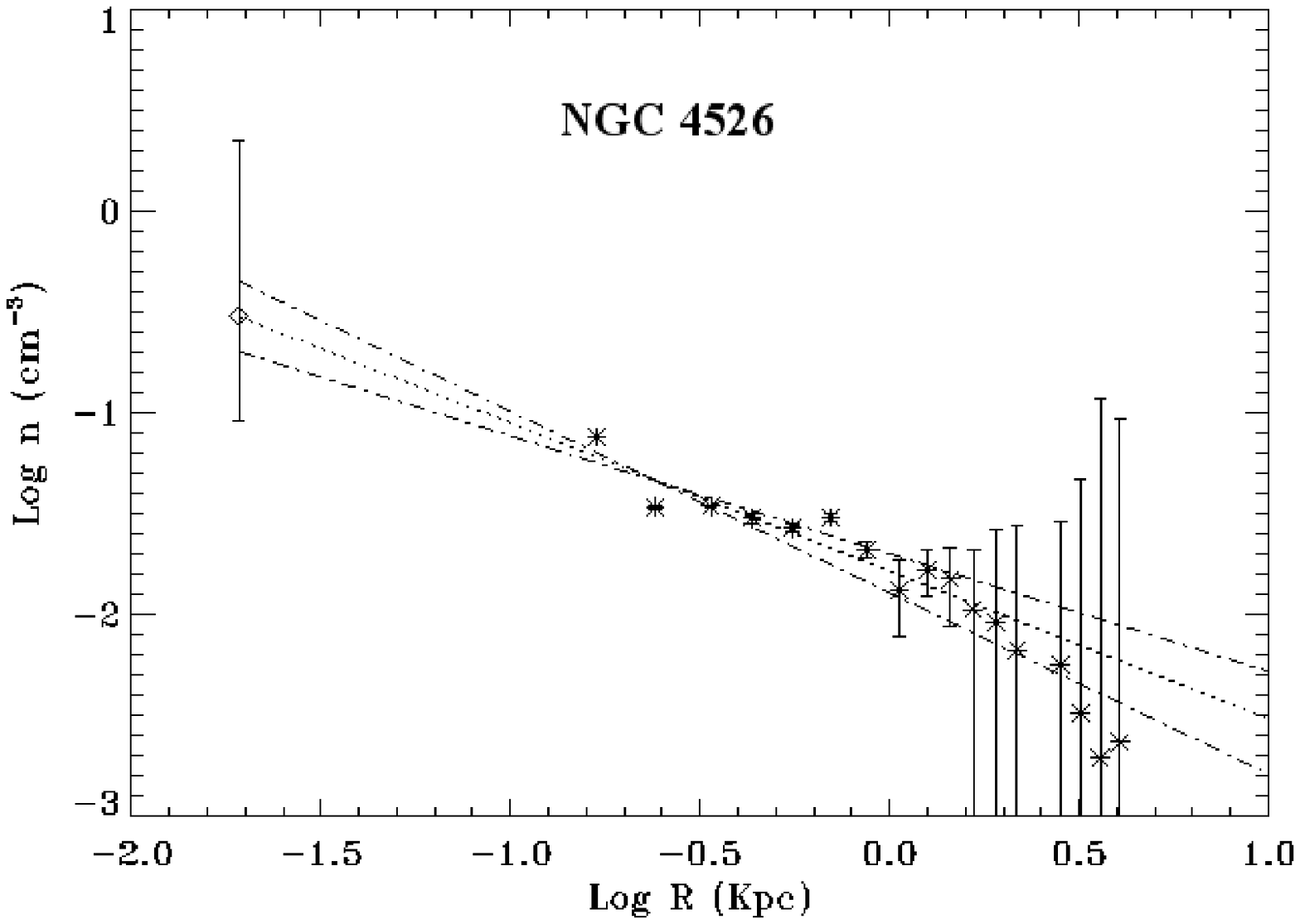,width=0.45\linewidth}}
\caption{Deprojected temperature (left panels) and density (right panels)
  profiles of the four VCC galaxies. The two dotted lines in each figure are
  the linear fits on the innermost data points, used for the extrapolation of
  the density to the Bondi radius.}
\label{figa2}
\end{figure*}

The second contaminating component originates from the stellar coronae of CVs
and ABs; their contribution cannot be separated through a spectral analysis as
for LMXB, because their thermal X-ray emission is similar to that of the ISM
(e.g. \citealt{revnivtsev07}).  We exploit conversely the result obtained by
\citet{revnivtsev08}: they show that the old stars contribution to the X-ray
emission of early-type galaxies (per unit stellar luminosity in the K band)
displays a small scatter (less than a factor of 2) around the value they
measured for NGC~3379, 
\begin{equation}
\frac{L_{X, \, 0.5-2 {\rm keV}}}{L_K} =
(6.9 \pm 0.7)\times10^{27} {\rm erg} \,{\rm s}^{-1}L^{-1}_{K,\odot}.
\end{equation}

We then retrieved the K band surface brightness profiles of the four galaxies
of interest from the 2MASS Large Galaxy Atlas. Assuming that the IR-to-X-ray
scaling relationship  also holds in our sources, we compared them with the
X-ray brightness profiles obtained from our analysis. In Fig. \ref{figa1b} we
show as an example the case of NGC~4365. This is the galaxy where we measured
the largest (possible) contribution from old stars that amounts to $\sim$ 30
\% of the total counts. We conclude that the diffuse soft X-ray light is
dominated by the genuine emission from the galaxies hot coronae.  Nonetheless,
we corrected the observed profiles for this marginal contamination.

After removing the contribution from stellar sources, we followed the same
procedure to derive the deprojected profiles of temperature and density as in
BBC08, to which we refer the reader for further details. Very briefly, to
measure the dependence of temperature with radius, we de-project the spectra
assuming spherical symmetry using the {\it PROJECT} model in XSPEC,
considering four annuli. The results are graphically reported in
Fig. \ref{figa2}. Concerning the density, we first of all determined the X-ray
surface brightness profile (SBP) extracting the counts in a series of circular
annuli.  The next step is to de-project the observed SBP, i.e. obtaining the
number of counts emitted per unit volume as a function of radius. Assuming
spherical symmetry, the count contribution provided by each spherical shell
to the inner ones were determined following the calculation by
\citet{kriss83}. From the deprojected SBP we obtained $n(r)$ assuming thermal
emission and inverting the normalization coefficient of the spectral model.

As the Bondi radius is well inside the innermost annulus (by a factor between
3 and 20), the value of the density was extrapolated down to $r_B$ through a
fit across the profile, assuming a power-law in the form $n(r) \propto
r^{-\alpha}$. To limit systematic errors related to the arbitrary choice of
the range of radii to be included in the fit, we performed the analysis twice
by using four and six of the innermost density points. For each case we
estimated the uncertainty related to the accuracy of the parameters of the fit
describing the density behavior.  We forced the density at $r_B$ to be higher
(or equal) than its measured value at the innermost annulus. We finally
adopted for $n_B$ the average of the values found in these two extrapolations
and for its uncertainty the full range given by the overlap of the two
individual error bars.

\end{document}